\documentclass[12pt]{iopart}
\usepackage{iopams}  
\usepackage{graphicx,psfrag,bbm,latexsym,color,dcolumn,bm,dsfont,bbm,color}

\newcommand{\ua}{\uparrow}
\newcommand{\da}{\downarrow}
\newcommand{\media}[1]{\langle #1 \rangle}
\newcommand{\E}{\mbox{$\mathsf E$}}

\begin{document}

\title[Ground state of many-body lattice systems:
an analytical probabilistic approach]
{Ground state of many-body lattice systems:\\
an analytical probabilistic approach}

\author{Massimo Ostilli$^{1,2}$\ and Carlo Presilla$^{1,2,3}$}

\address{$^1$\ Dipartimento di Fisica, Universit\`a di Roma ``La Sapienza'',
Piazzale A. Moro 2, Roma 00185, Italy}
\address{$^2$\ Center for Statistical Mechanics and Complexity, Istituto Nazionale per la Fisica della Materia, Unit\`a di Roma 1, Roma 00185, Italy}
\address{$^3$\ Istituto Nazionale di Fisica Nucleare, Sezione di Roma 1, 
Roma 00185, Italy}

\date{\today}

\begin{abstract}
On the grounds of a Feynman-Kac--type formula for Hamiltonian lattice systems 
we derive analytical expressions for the matrix elements of the 
evolution operator. 
These expressions are valid at long times when a central limit
theorem applies.
As a remarkable result we find that the ground-state energy 
as well as all the correlation functions in the ground state 
are determined semi-analytically by solving a simple scalar equation.
Furthermore, explicit solutions of this equation are obtained in the 
noninteracting case.
\end{abstract}

 \maketitle

\section{Introduction}

The Feynman-Kac formula \cite{SIMON} provides a powerful connection
between the imaginary time evolution of quantum systems in the continuum 
and probabilistic expectations over Wiener classical trajectories.  
Its extension to the case of multicomponent wavefunctions 
requires the introduction of a further expectation over stochastic 
Poisson processes \cite{DAJLS}.
The role of Poisson processes is central in the 
probabilistic representation of Berezin integrals over 
anticommuting variables and, in general, 
of the time evolution of discrete systems \cite{DJS}. 
In the simplified formulation \cite{PRESILLA} it is shown that
the real time or the imaginary time dynamics 
of systems described by a finite Hamiltonian matrix, representing
bosonic or fermionic degrees of freedom, 
is expressed in terms of the evolution of a proper collection 
of independent Poisson processes.
For a lattice system the Poisson processes are associated 
to the links of the lattice and their jump rates can be arbitrary.
In \cite{PRESILLA} it is demonstrated that when the rates of the 
Poisson processes are chosen equal to the hopping coefficients 
of the system the probabilistic representation leads to an optimal 
algorithm which coincides with the Green Function
Quantum Monte Carlo (QMC) method 
\cite{TRIVEDICEPERLEY,LINDEN,CALANDRASORELLA}
in the limit when the latter becomes exact \cite{SORELLACAPRIOTTI}.

In this paper we exploit the probabilistic representation 
\cite{PRESILLA} to derive analytical expressions for the matrix
elements of the evolution operator in the long time limit.
Our approach is based on a series expansion of the probabilistic
expectation in terms of conditional expectations with a fixed number 
$N$ of jumps of the Poisson processes. 
This resembles the expansion of the grand canonical partition function 
in statistical mechanics in terms of canonical averages with a fixed 
number of particles.
By integrating out the $N$ stochastic jump times,
we show that the conditional expectations become averages of functions
which depend only on the multiplicities $N_V$ and $N_A$ of the values assumed
by the potential and hopping energies, respectively, of the configurations
visited by the system.
According to a central limit theorem, 
at large values of $N$, \textit{i.e.} long times,
the rescaled multiplicities $N_V/\sqrt{N}$ and 
$N_A/\sqrt{N}$ become Gaussian distributed and
the corresponding averages can be evaluated analytically.
The parameters of the Gaussian probability density 
do not depend on the Hamiltonian parameters 
and are easily determined statistically.
Finally, the series of the conditional expectations can be 
re-summed with a saddle-point method. 
As a remarkable result, we find that the ground 
state energy is semi-analytically determined by solving a simple scalar
equation. 
Once this equation is solved the quantum expectation in the ground state 
for other operators can be determined analytically 
by using the Hellman-Feynman theorem.
The result is valid for boson or fermion systems. 
As regards fermions the well known sign problem 
\cite{GUBERNATIS,GUBERNATIS2,SORELLACAPRIOTTI,LEEUWEN0,WIESE,BECCARIA2}
is avoided by introducing an approximation related 
to the exact counting of the positive and negative contributions 
in the noninteracting case. 

The paper is organized as follows. 
In Section \ref{representation} we review the probabilistic representation
of quantum dynamics in the case of a generalized Hubbard Hamiltonian.
In Section \ref{decomposition} we decompose the expectation in 
canonical averages of weights which are calculated analytically.
The canonical averages are evaluated at long times 
via a central limit theorem in Section \ref{centrallimit}.
The equation for the ground-state energy is discussed in Section \ref{bosons}
for hard-core bosons and in Section \ref{fermions} for fermions.
In Section \ref{correlations} we show
how to calculate ground-state correlation functions
within our approach.
Finally in Section \ref{numericalresults} we show some example cases
and compare with the results of exact numerical calculations.
General features of our approach are summarized and discussed in
Section \ref{conclusions}.

\section{Exact probabilistic representation of lattice dynamics}
\label{representation}

We illustrate our approach in the case of imaginary time dynamics 
for a system of hard-core bosons or fermions described by
a generalized Hubbard Hamiltonian
\begin{eqnarray}\fl
\label{Hubbard}
H &=& - 
\sum_{i \neq j \in \Lambda} 
\sum_{\sigma=\ua\da} \eta_{ij}
c^\dag_{i\sigma} c^{}_{j\sigma} +
\sum_{i \in \Lambda} \gamma_i 
c^\dag_{i\ua} c^{}_{i\ua} c^\dag_{i\da} c^{}_{i\da}+
\sum_{i \in \Lambda} 
\sum_{\sigma=\ua\da} \delta_{i\sigma}
c^\dag_{i\sigma} c^{}_{i\sigma},
\end{eqnarray}
where $\Lambda\subset Z^d$ is a finite $d$-dimensional lattice
with $|\Lambda|$ sites
and $c_{i \sigma}$ the commuting or anticommuting 
destruction operators at site $i$ and spin index $\sigma$
with the property $c_{i\sigma}^2=0$.
We are interested in evaluating the matrix elements
$\langle \bm{n} |e^{- Ht} | \bm{n}_0 \rangle$
where $\bm{n}= (n_{1 \ua},n_{1 \da}, \ldots, n_{|\Lambda| \ua},
n_{|\Lambda| \da})$ are the lattice occupation numbers 
taking the values 0 or 1. 
The total number of particles is
$N_\sigma = \sum_{i \in \Lambda} n_{i\sigma}$ for $\sigma=\ua \da$.
In the following we shall use the mod 2 addition
$n \oplus n'= (n+n') \bmod 2$.

Let $\Gamma$ be the set of system links, \textit{i.e.} 
the unordered pairs $(i,j)$ with  
$i,j\in\Lambda$ such that  $\eta_{ij}\neq 0$. 
For simplicity, we will start by assuming $\eta_{ij}=\epsilon$  
if $i$ and $j$ are first neighbors and $\eta_{ij}=0$ otherwise.
We will also assume $\gamma_i=\gamma$ and $\delta_i=0$.
We shall call such a model the first neighbor uniform (FNU) model.
For a $d$-dimensional lattice the number of links per spin component is 
$|\Gamma| = d |\Lambda|$.
Let us introduce
\begin{eqnarray}
\label{lambda}
\lambda_{ij \sigma}(\bm{n}) &=& 
\langle \bm{n} \oplus \bm{1}_{i\sigma} \oplus \bm{1}_{j\sigma}|
c^\dag_{i\sigma} c^{}_{j\sigma} + c^\dag_{j\sigma} c^{}_{i\sigma}
|\bm{n}\rangle 
\\
\label{potential}
V(\bm{n}) &=& 
\langle \bm{n}|H|\bm{n}\rangle ,
\end{eqnarray}
where $\bm{1}_{i\sigma}=(0,\ldots,0,1_{i\sigma},0,\ldots,0)$, 
and let $\{N_{ij\sigma}^t\}$, $(i,j) \in \Gamma$, be a family
of $2|\Gamma|$ independent Poisson processes with jump rate $\rho$.
At each jump of the process $N_{ij\sigma}^t$, 
if $\lambda_{ij\sigma} \neq 0$ a particle moves from site $i$ to site $j$
or vice versa, while the lattice configuration $\bm{n}$ remains 
unchanged if $\lambda_{ij\sigma} = 0$.
The total number of jumps at time $t$ is 
$N_{t} = \sum_{(i,j)\in \Gamma,\sigma=\ua\da} N^t_{ij \sigma}$.
By ordering the jumps according to the times $s_{k}$, $k=1,\dots, N_{t}$, 
at which they take place in the interval $[0,t)$,
we define a trajectory as the Markov chain
$\bm{n}_{1}, \bm{n}_2, \dots, \bm{n}_{N_{t}}$ 
generated from the initial configuration $\bm{n}_0$.
Let us call 
$\lambda_{1},\lambda_{2}, \dots, \lambda_{N_{t}}$
and $V_{1},V_2 \dots, V_{N_{t}}$ the values of 
the matrix elements (\ref{lambda}) and (\ref{potential})
occurring along the trajectory.
As proved in \cite{PRESILLA}, the following representation holds
\begin{eqnarray}
\label{TheFormulaa}
\langle \bm{n}|e^{-Ht} | \bm{n}_0\rangle &=&  \E  \left(
\delta_{ \bm{n} , \bm{n}_{N_t}} 
{\cal M}^t \right), 
\end{eqnarray}
where the stochastic functional ${\cal M}^t$ is defined by
\begin{eqnarray}
\label{FORMULA D}
{\cal M}^t = e^{2|\Gamma|\rho t}  
\left( 
\prod_{k=1}^{N_{t}} \frac{\epsilon}{\rho} \lambda_k
e^{-V_{k-1}(s_{k}-s_{k-1})} 
\right) 
e^{-V_{N_{t}}(t-s_{N_{t}})}
\end{eqnarray}
if $N_t > 0$ and 
${\cal M}^t = e^{2|\Gamma|\rho t}  e^{-V_0 t}$ if $N_t=0$.
Here $V_0=V(\bm{n}_0)$ and $s_0=0$.

Several quantities can be obtained from the matrix 
elements (\ref{TheFormulaa}).
The ground-state energy is given by
\begin{equation}
\label{E0} 
E_0 = \lim_{t \to \infty}
\frac{-\sum_{\bm{n}} \partial_t \langle \bm{n}|e^{-Ht} | \bm{n}_0\rangle}
{\sum_{\bm{n}} \langle \bm{n}|e^{-Ht} | \bm{n}_0\rangle }
= \lim_{t \to \infty} 
\frac{- \partial_t \E ({\cal M}^t)}
{\E ({\cal M}^t)}.
\end{equation}

\section{Canonical decomposition of expectation}
\label{decomposition}

To evaluate (\ref{E0}) we decompose the expectation 
$\E ({\cal M}^{t})$ as a series of conditional 
expectations with a fixed number of jumps (canonical averages)
\begin{eqnarray}
\label{EXPANSION}
\E ({\cal M}^{t}) &=&
\sum_{N=0}^{\infty}\E ({\cal M}^{t}|N_{t}=N) 
\nonumber\\ &=&
\sum_{N=0}^{\infty}  
\sum_{r \in \Omega_N}
\mathcal{S}_N^{(r)} \mathcal{W}_N^{(r)}(t),
\end{eqnarray}
where $\Omega_N=\Omega_N(\bm{n}_{0})$ is the set of trajectories 
with $N$ jumps branching from the initial configuration $\bm{n}_{0}$
and 
\begin{eqnarray}
\label{signs}
\mathcal{S}_N^{(r)} =
\lambda_1^{(r)} \lambda_2^{(r)} \dots \lambda_N^{(r)} ,
\end{eqnarray} 
\begin{eqnarray}
\label{weights}
\fl \mathcal{W}_N^{(r)}(t) =
\epsilon^{N} \int_{0}^{t}ds_{1} 
\int_{s_{1}}^{t}ds_{2} 
\dots 
\int_{s_{N-1}}^{t}ds_{N}
e^{-V_{0}s_{1}
-V_1^{(r)}(s_{2}-s_{1})
- \dots 
-V_N^{(r)}(t-s_{N})}.
\end{eqnarray} 
The weights (\ref{weights}) are obtained on multiplying (\ref{FORMULA D})
by the infinitesimal probability 
$e^{-2|\Gamma| \rho t} \rho^N ds_{1} ds_{2} \dots ds_{N}$
to have $N$ jumps and integrating over the jump times.

A link $ij$ with spin $\sigma$ is called active if
$\lambda_{ij\sigma}\neq 0$.
From (\ref{signs}) it is clear that only trajectories
formed by a sequence of active links contribute to (\ref{EXPANSION}). 
Hereafter we restrict $\Omega_N$ 
to be the set of these effective trajectories with $N$ jumps.
The sum over the set $\Omega_N$ in (\ref{EXPANSION}) can be
rewritten as an average, $\media{\cdot}$, 
over the trajectories with $N$ jumps generated
by extracting with uniform probability one of the active links 
available at the configurations 
$\bm{n}_0,\bm{n}_1,\ldots,\bm{n}_{N-1}$.
If $A_k = \sum_{(i,j)\in \Gamma,\sigma=\ua\da} 
|\lambda_{ij\sigma}(\bm{n}_k)|$ is the number of active links 
in the configuration $\bm{n}_k$, the probability associated 
to the trajectory $r$ is 
$p_N^{(r)} = \prod_{k=0}^{N-1} 1/{A_k^{(r)}}$ and we have
\begin{eqnarray}
\label{AVERAGE}
\sum_{r \in \Omega_N} \mathcal{S}_N^{(r)} \mathcal{W}_N^{(r)}(t)
&=&
\sum_{r \in \Omega_N} 
p_N^{(r)} \mathcal{S}_N^{(r)} \mathcal{W}_N^{(r)}(t)
\prod_{k=0}^{N-1} A_k^{(r)}
\nonumber\\&=&
\media{
\mathcal{S}_N
\mathcal{W}_N(t)
\prod_{k=0}^{N-1} A_k}.
\end{eqnarray}
Note that $\prod_{k=0}^{N-1} A_k = \prod_A A^{N_A}$
depends only on the multiplicities $N_A$ of the 
values $A$ assumed by the number of active links;
these multiplicities are normalized to $N$, 
\textit{i.e.} $\sum_A N_A = N$.
For the FNU model the possible values of $A$ 
depend on the number of particles
and we have the bound $A\leq \min(2d(N_\ua + N_\da), 2|\Gamma|)$. 

For a generic trajectory the weights (\ref{weights}) satisfy the 
recursive differential equation 
\begin{eqnarray}
\frac{d \mathcal{W}_N(t)}{dt} =
\mathcal{W}_{N-1}(t) -V_N \mathcal{W}_N(t),
\end{eqnarray}
where $\mathcal{W}_{-1}(t) = 0$.
In terms of the Laplace transform
$ \widetilde{\mathcal{W}}_{N}(z)
= \int_{0}^{\infty}dt e^{-zt} \mathcal{W}_{N}(t) $
we get
\begin{eqnarray}
\label{WTILDE}
\widetilde{\mathcal{W}}_{N}(z) = 
\epsilon^{N} \prod_{k=0}^{N}\frac{1}{z+V_{k}}
= \epsilon^{N} \prod_{V}\frac{1}{(z+V)^{N_V}}.
\end{eqnarray}
In (\ref{WTILDE}) we see that
the weights depend only on the multiplicities $N_V$ of the 
values $V$ assumed by the potential;
these multiplicities are normalized to $N+1$, 
\textit{i.e.} $\sum_V N_V = N+1$.
For the model (\ref{Hubbard}) the possible values assumed 
by $V$ are $V=0,\gamma,2\gamma,\ldots,N_p\gamma$, 
where $N_p=\min(N_\ua,N_\da)$.

For an assigned set of multiplicities of the potential, 
the antitransform of (\ref{WTILDE}) can be evaluated 
with the residue method.
In this way one obtains an exact recursive expression of
$\mathcal{W}_N(t)$
which, however, for large $N$ must be evaluated numerically 
with multi-precision algebra.
On the other hand, for $N$ large a complex saddle-point method 
can be used which provides the following asymptotically exact 
explicit expression
\begin{eqnarray}
\label{WSADDLE}
\mathcal{W}_{N}(t) =
\frac{e^{x_{0}t-\sum_{V}N_{V}\log[(x_{0}+V)/\epsilon]}}
{\sqrt{ 2\pi \sum_{V}\frac{\epsilon^2 N_{V}}{(x_{0}+V)^{2}} } },
\end{eqnarray}
where $x_{0}$ is the solution of the equation
\begin{eqnarray}
\label{X}
\sum_{V}\frac{N_{V}}{x_{0}+V}=t.
\end{eqnarray}
Note that in the case $\gamma=0$, Eq. (\ref{WSADDLE})
reduces to Stirling's approximation of the exact value 
$\mathcal{W}_{N}(t) = \epsilon^N t^N/N!$.

\section{Canonical averages via a central limit theorem}
\label{centrallimit}

As is evident from the explicit expression given 
in the case $\gamma=0$,
the weights $\mathcal{W}_{N}(t)$ have a maximum at some $N$ which 
increases by increasing $t$. 
This remains true also in the general case, as shown in the following.
Therefore in the long time limit the most important contributions
to the expansion (\ref{EXPANSION}) of the expectation $\E ({\cal M}^{t})$ 
come from larger and larger values of $N$.
In this section we will evaluate the canonical averages (\ref{AVERAGE})
analytically for $N$ large by using the asymptotic behavior of
the stochastic variables $N_V$ and $N_A$.
In this limit we will not distinguish the different normalizations,
$N+1$ and $N$, of $N_V$ and $N_A$, respectively.
In the following, we will indicate with $m_V$ and $m_A$ the number
of different values assumed by the variables $V$ and $A$, respectively.
For clarity, we consider separately the hard-core boson and fermion cases. 

\subsection{Hard-core bosons}
\label{bosons}
In this case we have $\mathcal{S}_N = 1$
and the canonical averages (\ref{AVERAGE}) are averages 
of a function which depends only on the multiplicities, 
$N_V$ and $N_A$ (besides a parametric dependence on time). 
In terms of the corresponding frequencies, 
$\nu_V=N_V/N$ and $\nu_A=N_A/N$, 
which for $N$ large become continuously distributed in the range
$[0,1]$ with the constraints 
\begin{eqnarray}
\label{CONSTRAINTS}
\sum_V \nu_V = \sum_A \nu_A =1,
\end{eqnarray}
Eq. (\ref{AVERAGE}) can be rewritten as
\begin{eqnarray}
\label{AVERAGE1}
\media{ \mathcal{W}_{N}(t) \! \prod_{k=0}^{N-1} \! A_{k} } =
\int d\bm{\nu}
\mathcal{P}_N(\bm{\nu}) 
g_N(t;\bm{\nu})
\\ 
g_N(t;\bm{\nu}) =
\frac{e^{x_{0}t+ N \left( \bm{\nu},\bm{u} \right) }}
{ \sqrt {2\pi N 
\sum_{V}\frac{\epsilon^2 \nu_{V}}{(x_{0}+V)^{2}} }} ,
\label{G}
\end{eqnarray}
where $\bm{\nu}$ and $\bm{u}$ are vectors with
$m=m_V+m_A$ components defined as
$\bm{\nu}^T = (\ldots \nu_V \ldots; \ldots \nu_A \ldots)$ and
$\bm{u}^T= (\ldots -\log[(x_0+V)/\epsilon] \ldots;\ldots \log A \ldots)$,
respectively.
Note that $\bm{u}$ depends on $\bm{\nu}$ through $x_{0}=x_{0}(\bm{\nu})$.

The probability density $\mathcal{P}_N(\bm{\nu})$ is
given by the fraction of trajectories 
branching from the initial configuration $\bm{n}_0$ and having
after $N$ jumps multiplicities $N_V=\nu_V N$ and $N_A=\nu_A N$.
For $N$ large, it can be approximated in the following way. 
We rewrite the multiplicities as 
$N_V = \sum_{k=0}^N \chi_V(\bm{n}_k)$ and 
$N_A = \sum_{k=1}^N \chi_A(\bm{n}_{k-1})$,
where $\chi_V(\bm{n})=1$ if $V(\bm{n})=V$ and $\chi_V(\bm{n})=0$ 
otherwise, and similarly for $\chi_A$.
Since the configurations $\bm{n}_k$ form a Markov chain
with finite state space, a central limit theorem applies to
each rescaled sum $N_V/\sqrt{N}$ or $N_A/\sqrt{N}$
\cite{BILLINGSLEY}.
However, due to the constraints (\ref{CONSTRAINTS}), 
the joint probability for these $m$ rescaled sums 
is not Gaussian. 
Given an arbitrary set of $m_V-1$ 
$V$-like components and $m_A-1$ $A$-like components, 
the joint probability density is the product 
of a Gaussian density for this set of variables
and two delta functions which take into account the constraints
(\ref{CONSTRAINTS}).
For the frequencies $\bm{\nu}$, therefore, we have 
\begin{eqnarray}
\label{DENSITY}
\mathcal{P}_N(\bm{\nu}) = \mathcal{F}_N(\hat{\bm{\nu}}) 
~\delta \Big( \sum_V \nu_V-1 \Big) 
~\delta \Big( \sum_A \nu_A-1 \Big),
\end{eqnarray}
where $\mathcal{F}_N(\hat{\bm{\nu}})$ is the normal density defined
in terms of the vector $\hat{\bm{\nu}}$ having the 
$m-2$ chosen components of $\bm{\nu}$
\begin{eqnarray}
\label{GAUSSIAN}
\mathcal{F}_N(\hat{\bm{\nu}}) =
\sqrt{ 
\frac{N^{m-2} |\det \hat{\bm{\Sigma}}^{-1}|} {(2\pi)^{m-2}} }
~e^{ -\frac{N}{2} 
\left( \hat{\bm{\Sigma}}^{-1} (\hat{\bm{\nu}} - 
\hat{\overline{\bm{\nu}}}),
(\hat{\bm{\nu}} - \hat{\overline{\bm{\nu}}}) \right) }.
\end{eqnarray}
Here
$\hat{\overline{\bm{\nu}}}$ and $\hat{\bm{\Sigma}} N^{-1}$ 
are the $(m-2)$-component subvector and submatrix, respectively,  
of the mean value, $\overline{\bm{\nu}}$, and
the covariance matrix, $\bm{\Sigma} N^{-1}$, of $\bm{\nu}$.
As discussed in Section \ref{numericalresults},
the quantities $\overline{\bm{\nu}}$ and $\bm{\Sigma}$ are easily 
measured by sampling over trajectories with a large number of jumps.

By using (\ref{DENSITY}), the $m$-dimensional integral which appears
in (\ref{AVERAGE1}) can be performed by the saddle-point method. 
Note that this integration method is asymptotically exact for $N$ large.
Due to the constraints (\ref{CONSTRAINTS}),
$\overline{\bm{\nu}}$ satisfies the property 
$\sum_V \overline{\nu}_V = \sum_A \overline{\nu}_A =1$
while rows and columns of the $VV$, $VA$, $AV$, and $AA$ blocks 
of $\bm{\Sigma}$ are normalized to zero. 
By using these properties, in terms of $\overline{\bm{\nu}}$ 
and $\bm{\Sigma}$ we get
\begin{eqnarray}
\label{AVERAGE2}
\left.
\media{ \mathcal{W}_{N}(t) \! \prod_{k=0}^{N-1} \! A_{k} } =
\frac{e^{x_{0}t + N \left[
(\overline{\bm{\nu}},\bm{u}) 
+ \frac12 (\bm{\Sigma} \bm{u},\bm{u}) \right] }}
{ \sqrt{2\pi N 
\sum_{V}\frac{\epsilon^2 \nu_V}{(x_{0}+V)^{2}}  }} 
\right|_{\bm{\nu}=\bm{\nu}^\mathrm{sp}},
\end{eqnarray}
where 
$\bm{\nu}^\mathrm{sp}$ is the saddle-point frequency defined
by the equation
\begin{eqnarray}
\label{saddle}
\bm{\nu}^\mathrm{sp} = \overline{\bm{\nu}} + 
\bm{\Sigma} \bm{u}(\bm{\nu}^\mathrm{sp}).
\end{eqnarray}

In order to evaluate the expectation value $\E ({\cal M}^{t})$ 
we need to sum the series (\ref{EXPANSION}).
This can be done with a further saddle-point integration. 
According to Eq. (\ref{AVERAGE2}) 
the terms of this series are exponentially peaked at 
$N=N^\mathrm{sp}$, where $N^\mathrm{sp}$ satisfies
\begin{eqnarray}\fl
\label{Nsaddle}
\left[ \left\{ t 
-N \left[ ( \overline{\bm{\nu}},\bm{v} ) +
\left( \bm{\Sigma} \bm{u}, \bm{v} \right) \right]
 \right\}
\frac {\partial x_{0}}{\partial N}+
 (\overline{\bm{\nu}}{},{\bm{u}}) + \frac{1}{2} 
\left( \bm{\Sigma} {\bm{u}}, {\bm{u}} \right)
\right]_{\bm{\nu}=\bm{\nu}^\mathrm{sp},{N=N^\mathrm{sp}}}
=0,
\end{eqnarray}
with
$\bm{v}^T = (\ldots (x_0+V)^{-1} \ldots; \ldots 0 \ldots)$.
We observe that, according to Eqs. (\ref{X}) and (\ref{saddle}),
the term 
$\left\{ t -N \left[ ( \overline{\bm{\nu}},\bm{v} ) +
\left( \bm{\Sigma} \bm{u}, \bm{v} \right) \right] \right\}$ 
vanishes for $\bm{\nu}=\bm{\nu}^\mathrm{sp}$ 
so that the above condition reduces to
\begin{eqnarray}
\label{E1}
\left[
(\overline{\bm{\nu}}{},{\bm{u}}) + \frac{1}{2} 
\left( \bm{\Sigma} {\bm{u}}, {\bm{u}} \right)
\right]_{\bm{\nu}=\bm{\nu}^\mathrm{sp},{N=N^\mathrm{sp}}}
=0.
\end{eqnarray}
Equation (\ref{E1}) is a time independent equation which determines
$\left. x_0 \right|_{\bm{\nu}=\bm{\nu}^\mathrm{sp},N=N^\mathrm{sp}}$
as a function of $\overline{\bm{\nu}}$ and  $\bm{\Sigma}$. 
According to Eq. (\ref{X}),
this means that for $\bm{\nu}=\bm{\nu}^\mathrm{sp}$, 
the quantity $N^\mathrm{sp}$ increases linearly with time so that 
$\left. x_0 \right|_{\bm{\nu}=\bm{\nu}^\mathrm{sp},N=N^\mathrm{sp}}$
becomes independent of time.
In conclusion, a saddle-point integration with respect to $N$ 
of the series (\ref{EXPANSION}) provides
\begin{eqnarray}
\E ({\cal M}^{t}) = 
\left. \frac{e^{x_0 t}}{ \sum_V \frac{\epsilon \nu_V}{x_0+V}}
\right|_{\bm{\nu}=\bm{\nu}^\mathrm{sp},N=N^\mathrm{sp}}.
\end{eqnarray} 
By taking the time derivative of this expectation and 
using (\ref{E0}), we obtain that the ground-state energy 
of the hard-core boson system is
\begin{eqnarray}
\label{E0B}
E_{0B} = 
- \left. x_0 \right|_{\bm{\nu}=\bm{\nu}^\mathrm{sp},N=N^\mathrm{sp}}.
\end{eqnarray}
Equation (\ref{E1}) is, therefore, the equation for the ground-state energy.
It defines $E_{0B}$ in terms of $\overline{\bm{\nu}}$
and $\bm{\Sigma}$ and explicitly reads 
\begin{eqnarray}
\label{E2}
0 &=&
-\sum_{V}\overline{{\nu}}_V\log\left(\frac{-E_{0B}+V}{\epsilon}\right)
+\sum_A\overline{{\nu}}_A\log\left(A\right)
\nonumber \\ &&
+\frac{1}{2}\sum_{V,V'} {\Sigma}_{V,V'}
\log\left(\frac{-E_{0B}+V}{\epsilon}\right)
\log\left(\frac{-E_{0B}+V'}{\epsilon}\right)  
\nonumber \\ &&
-\sum_{V,A} \Sigma_{V,A}
\log\left(\frac{-E_{0B}+V}{\epsilon}\right) \log\left(A\right)
\nonumber \\ &&
+\frac{1}{2}\sum_{A,A'}{\Sigma}_{A,A'}
\log\left(A\right)\log\left(A'\right) .
\end{eqnarray}
In the case $\gamma=0$ the ground-state energy $E_{0B}^{(0)}$ 
can be solved explicitly and one has
\begin{eqnarray}
\label{E_{0B}0}
E_{0B}^{(0)}=-\epsilon
\exp\left[
\sum_A\overline{{\nu}}_A\log(A)+
\frac{1}{2}\sum_{A,A'} {\Sigma}_{A,A'} \log(A)\log(A') 
\right]. 
\end{eqnarray}
Note that Eq. (\ref{E_{0B}0}) is a nontrivial 
formula for the ground state of a system of bosons interacting via
a hard-core potential.
With the above expression Eq. (\ref{E2}) can be written more compactly as   
\begin{eqnarray}
\label{E3}
 \log\left(\frac{-E_{0B}^{(0)}}{\epsilon}\right) 
&=&
\sum_{V}\overline{{\nu}}_V\log\left(\frac{-E_{0B}+V}{\epsilon}\right)
\nonumber \\ &&
-\frac{1}{2} \sum_{V,V'} {\Sigma}_{V,V'}
\log\left(\frac{-E_{0B}+V}{\epsilon}\right)
\log\left(\frac{-E_{0B}+V'}{\epsilon}\right)
\nonumber \\ &&
+\sum_{V,A}\Sigma_{V,A}
\log\left(\frac{-E_{0B}+V}{\epsilon}\right)\log\left(A\right) .
\end{eqnarray}
By using the bounds $E_{0B}^{(0)} <E_{0B} <0$,
the scalar Eq. (\ref{E3}) can be easily solved 
with the bisection method.

The generalization of the above results to Hubbard Hamiltonians
(\ref{Hubbard})
with arbitrary parameters $\bm{\eta},\bm{\gamma},\bm{\delta}$ 
is straightforward.
Equations (\ref{AVERAGE2}-\ref{E0B}) remain formally unchanged,
however the vectors $\overline{\bm{\nu}},\bm{u},\bm{v}$ and the
covariance matrix $\bm{\Sigma}$ are modified in order to  
take into account all the possible values of the generalized 
potential $V$ corresponding to the operator 
$\sum_{i \in \Lambda} \gamma_i 
c^\dag_{i\ua} c^{}_{i\ua}~c^\dag_{i\da} c^{}_{i\da}+
\sum_{i \in \Lambda} 
\sum_{\sigma=\ua\da} \delta_{i\sigma}
c^\dag_{i\sigma} c^{}_{i\sigma}$
and all the possible values of the generalized kinetic quantities
$T=A \eta/\epsilon$, where now $\epsilon$ is a unity of energy
and $\eta/\epsilon$ is the dimensionless hopping
value corresponding to the current jumping link. 
In fact the generalization of Eq. (\ref{AVERAGE}) 
consists in replacing $\prod_{k=0}^{N-1}A_k$
with $\prod_{k=0}^{N-1}(A_k\eta_k/\epsilon)$. 
Explicitly, 
now the vectors $\overline{\bm{\nu}}$ and $\bm{u}$ are
\begin{eqnarray}
\overline{\bm{\nu}}^T = (\ldots \overline{\nu}_V \ldots; \ldots \overline{\nu}_T \ldots) \nonumber \\
\bm{u}^T= 
(\ldots -\log[(x_0+V)/\epsilon] \ldots;\ldots \log T \ldots)
\end{eqnarray}
and  
the generalized ground-state energy $E_{0B}^{(0)}$ 
corresponding to $\bm{\gamma}=\bm{\delta}=\bm{0}$ reads
\begin{eqnarray}
\label{E_{0B}0G}
E_{0B}^{(0)}=-\epsilon
\exp\left[
\sum_T\overline{{\nu}}_T\log(T)+
\frac{1}{2}\sum_{T,T'} {\Sigma}_{T,T'} \log(T)\log(T') 
\right].
\end{eqnarray}

Similar considerations hold in the case of Hamiltonians
with arbitrary potential operators.

\subsection{Fermions}
\label{fermions}
In this case we have $\mathcal{S}_N = \pm 1$ and 
the canonical averages (\ref{AVERAGE}) are averages 
of a function which depends not only on the multiplicities, 
$N_V$ and $N_A$ but also on $N_{-}$, the sign multiplicity
related to $\mathcal{S}_N$ by $\mathcal{S}_N = (-1)^{N_{-}}$.
The approach developed in the case of hard-core bosons can be extended
also to fermions by including this further multiplicity $N_{-}$.
We will report on this procedure elsewhere. 
Here, we introduce an approximation which allows to reduce
the calculation of the fermion ground-state energy to that
of an effectively modified hard-core boson system.
This approximation is motivated by the 
observation that the correlations between $N_-$ and $N_V$ are smaller
than those between $N_-$ and $N_A$.

Let us consider again the FNU model.
In order to evaluate (\ref{AVERAGE}) for a fermion system we 
introduce the average weighted sign $s_N$ after $N$ jumps   
\begin{eqnarray}
\label{AVERAGEF}
s_N = 
\frac{\media{ \mathcal{S}_N \mathcal{W}_N(t) \prod_{k=0}^{N-1} A_k}}
{\media{ \mathcal{W}_N(t) \prod_{k=0}^{N-1} A_k}}.
\end{eqnarray}
The quantity $s_N$ is a function of the interaction strength.
For $\gamma=0$ it can be evaluated in the following way. 
Expanding $\sum_{\bm{n}} \langle \bm{n}|e^{-Ht} | \bm{n}_0\rangle$
in powers of $t$ 
and comparing with the expansion (\ref{EXPANSION}), 
for $\gamma=0$ and $N$ large we get in the case of hard-core bosons 
and fermions, respectively
\begin{eqnarray}
\media{ \prod_{k=0}^{N-1} A_k}
= c_{0B} (\bm{n}_0) \left( -E_{0B}^{(0)} / \epsilon \right)^N
\\
\media{ \mathcal{S}_N \prod_{k=0}^{N-1} A_k}
= c_{0F} (\bm{n}_0) \left( -E_{0F}^{(0)} / \epsilon \right)^N,
\end{eqnarray}
where $c_{0B} (\bm{n}_0)$ and $c_{0F} (\bm{n}_0)$ are coefficients 
related to the initial configuration $\bm{n}_0$ and 
$E_{0B}^{(0)}$ and $E_{0F}^{(0)}$ are the $\gamma=0$ ground-state 
energies.
The average weighted sign after $N$ jumps for $\gamma=0$
is then given by 
\begin{eqnarray}
\label{SN}
s_N  
= \frac{c_{0F} (\bm{n}_0)} {c_{0B} (\bm{n}_0)}~ 
e^{N \log \left( E_{0F}^{(0)}/E_{0B}^{(0)} \right)}.
\end{eqnarray}
In the case of fermions the noninteracting energy, $E_{0F}^{(0)}$, 
is known exactly while for hard-core bosons $E_{0B}^{(0)}$ 
can be computed with Monte Carlo 
simulations or analytically as shown above.
In general $E_{0F}^{(0)}/E_{0B}^{(0)}<1$ so that 
$s_N $ vanishes exponentially for $N$ large.

Approximating $s_N$ with its value (\ref{SN}) at $\gamma=0$
removes effectively the negative signs in the expectation
${\E ({\cal M}^t)}$. Therefore this can be evaluated 
as in the case of hard-core bosons with the same
Gaussian probability density. 
In particular the saddle-point condition for $N=N^{\mathrm{sp}}$ 
now becomes
\begin{eqnarray}
\label{E1F}
\left[
(\overline{\bm{\nu}}{},{\bm{u}}) + \frac{1}{2} 
\left( \bm{\Sigma} {\bm{u}}, {\bm{u}} \right)
+ \log \frac{E_{0F}^{(0)}}{E_{0B}^{(0)}}
\right]_{\bm{\nu}=\bm{\nu}^\mathrm{sp},{N=N^\mathrm{sp}}}
=0,
\end{eqnarray}
which is a time independent equation determining
$\left. x_0 \right|_{\bm{\nu}=\bm{\nu}^\mathrm{sp},N=N^\mathrm{sp}}$
in the fermion case. Finally, as in the hard-core boson case one finds that 
\begin{eqnarray}
\label{E0F}
E_{0F} = 
- \left. x_0 \right|_{\bm{\nu}=\bm{\nu}^\mathrm{sp},N=N^\mathrm{sp}},
\end{eqnarray}
and Eq. (\ref{E1F}) becomes equal to Eq. (\ref{E3})
with $E_{0B}$ and $E_{0B}^{(0)}$ substituted by
$E_{0F}$ and $E_{0F}^{(0)}$, respectively. However, in this case
we do not have the analogue of Eq. (\ref{E_{0B}0}) and 
$E_{0F}^{(0)}$ must be provided by an independent calculation,
\textit{i.e.} by exact diagonalization of the separable 
many-particle Hilbert space.

\section{Ground-state correlation functions}
\label{correlations}
The equation obtained in the previous Section for the determination
of the ground-state energy
depends on the parameters of the Hamiltonian only explicitly through the
values of the generalized potentials $V$ and of the kinetic quantities $T$.
In fact, the statistical moments $\overline{\bm{\nu}}$ and $\bm{\Sigma}$
are determined by the structure of the Hamiltonian not by the values
of the Hamiltonian parameters.
Therefore we are able to evaluate the derivatives of the
ground-state energy with respect to any parameter $\xi$
of the Hamiltonian $H(\xi)$.
This allows the determination of arbitrary ground-state correlation functions
via the Hellman-Feynman theorem
\begin{equation}
\label{HF}
\frac{\langle E_0(\xi) 
|\partial_{\xi} H(\xi)| 
E_0(\xi)\rangle}
{\langle E_0(\xi) |E_0(\xi)\rangle}=
\partial_{\xi} E_{0}(\xi), 
\end{equation}
where $E_0(\xi)$ is the ground-state energy of $H(\xi)$.
Hereafter, we assume a normalized ground state,
$\langle E_0(\xi) |E_0(\xi)\rangle=1$. 

Suppose that we want to evaluate the quantum expectation
of an operator $O$ in the ground state of the Hamiltonian $H$.
We have two possibilities. 

\textit{i)} The operator $O$ is a term
of the Hamiltonian itself, \textit{e.g.}
$O=c^\dag_{i\sigma} c^{}_{j\sigma}$, with $i\neq j$, 
$O=c^\dag_{i\ua} c^{}_{i\ua}~c^\dag_{i\da} c^{}_{i\da}$, or
$O=c^\dag_{i\sigma} c^{}_{i\sigma}$ 
if $H$ is the generalized Hamiltonian (\ref{Hubbard}).
In this case, by using Eq. (\ref{HF}) for hard-core bosons we have
\begin{eqnarray}
\label{O3}
\langle E_{0B}(\bm{\eta},\bm{\gamma},\bm{\delta}) |
c^\dag_{i\sigma} c^{}_{j\sigma}| 
E_{0B}(\bm{\eta},\bm{\gamma},\bm{\delta})
\rangle
&=&-\partial_{\eta_{ij}} E_{0B}(\bm{\eta},\bm{\gamma},\bm{\delta})
\nonumber\\&=&
\frac{\sum_{T}\frac{\nu_{T}^\mathrm{sp}}{T}
~\partial_{\eta_{ij}}T}
{\sum_{V}\frac{\nu_{V}^\mathrm{sp}}
{-E_{0B}(\bm{\eta},\bm{\gamma},\bm{\delta})+V}},
\qquad i\neq j
\end{eqnarray}
\begin{eqnarray}
\label{O1}
\langle E_{0B}(\bm{\eta},\bm{\gamma},\bm{\delta}) |
c^\dag_{i\ua} c^{}_{i\ua}~c^\dag_{i\da} c^{}_{i\da}| 
E_{0B}(\bm{\eta},\bm{\gamma},\bm{\delta})
\rangle
&=&\partial_{\gamma_i} E_{0B}(\bm{\eta},\bm{\gamma},\bm{\delta})
\nonumber\\&=&
\frac{\sum_{V}\frac{\nu_{V}^\mathrm{sp}}
{-E_{0B}(\bm{\eta},\bm{\gamma},\bm{\delta})+V}
~\partial_{\gamma_i} V}
{\sum_{V}\frac{\nu_{V}^\mathrm{sp}}
{-E_{0B}(\bm{\eta},\bm{\gamma},\bm{\delta})+V}},
\end{eqnarray}
\begin{eqnarray}
\label{O2}
\langle E_{0B}(\bm{\eta},\bm{\gamma},\bm{\delta}) 
|c^\dag_{i\sigma} c^{}_{i\sigma}| 
E_{0B}(\bm{\eta},\bm{\gamma},\bm{\delta})
\rangle
&=&\partial_{\delta_{i\sigma}} E_{0B}(\bm{\eta},\bm{\gamma},\bm{\delta})
\nonumber\\&=&
\frac{\sum_{V}\frac{\nu_{V}^\mathrm{sp}}
{-E_{0B}(\bm{\eta},\bm{\gamma},\bm{\delta})+V}
~\partial_{\delta_{i\sigma}} V}
{\sum_{V}\frac{\nu_{V}^\mathrm{sp}}
{-E_{0B}(\bm{\eta},\bm{\gamma},\bm{\delta})+V}}.
\end{eqnarray}
The expressions in the second lines of (\ref{O3}-\ref{O2}) 
have been obtained by using the derivatives of Eq. (\ref{E1}), 
which for a generic parameter $\xi$ read 
\begin{eqnarray}
\label{DE}
-\sum_V \nu^{\mathrm{sp}}_V \partial_{\xi}
\log\left(\frac{-E_{0B}+V}{\epsilon}\right)+
\sum_T \nu^{\mathrm{sp}}_T \partial_{\xi}\log\left(T \right)=0,
\end{eqnarray}
where $\bm{\nu}^\mathrm{sp}$ is given by (\ref{saddle}) 
and is determined once $E_{0B}(\bm{\eta},\bm{\gamma},\bm{\delta})$
has been solved.
Similar expressions hold in the case of fermions by 
using the derivatives of Eq. (\ref{E1F}) 
\begin{eqnarray}\fl
-\sum_V \nu^{\mathrm{sp}}_V \partial_{\xi}
\log\left(\frac{-E_{0F}+V}{\epsilon}\right)+
\sum_T \nu^{\mathrm{sp}}_T \partial_{\xi}\log\left(T \right)
+\partial_{\xi} \log\left( \frac{E_{0F}^{(0)}}{E_{0B}^{(0)}} \right)
=0.
\end{eqnarray}

\textit{ii)} If the operator $O$ is not a term of the Hamiltonian $H$
we consider a new Hamiltonian $H(\xi)=H+\xi O$ and 
calculate the corresponding ground-state energy $E_0(\xi)$.
Note that, since the used probabilistic representation holds for any system
described by a finite Hamiltonian matrix \cite{PRESILLA},  
the nature of the operator $O$ is arbitrary.
As an example we study the spin-spin structure factor
\begin{eqnarray}
\label{S}
S(q_x,q_y) =  \frac{1}{|\Lambda|}
\sum_{i,j \in \Lambda} 
e^{iq_x\left(x_i - x_j\right) + iq_y\left(y_i - y_j\right) }
\langle E_0 | S_i S_j  | E_0 \rangle,
\end{eqnarray}
where $S_i = c^\dag_{i\ua} c_{i\ua} - c^\dag_{i\da} c_{i\da}$ and
$x_i$ and $y_i$ are the coordinates of the $i$-th lattice point.
The quantum expectation of the operators $S_iS_j$ in the ground state
of the hard-core boson FNU model,  
$\langle E_{0B}(\epsilon,\gamma) 
| S_i S_j | E_{0B}(\epsilon,\gamma) \rangle$,
can be obtained by considering the Hamiltonians
\begin{eqnarray}
\label{Hij}
H(\xi_{ij})=H + \xi_{ij} S_i S_j,
\end{eqnarray}
where $H$ represents the FNU model.
For these Hamiltonians, the possible values of the potential are
$V=m\gamma+k\xi_{ij}$, with $m=0,1,\ldots,N_p$ and $k=-1,0,1$,  
and we have
\begin{eqnarray}\fl
\label{S1}
\langle E_{0B}(\epsilon,\gamma) 
| S_i S_j | 
E_{0B}(\epsilon,\gamma) \rangle=
\left.
\partial_{\xi_{ij}} E_{0B}(\epsilon,\gamma,\xi_{ij})
\right|_{\xi_{ij}=0}=
\left.
\frac{\sum_{V}\frac{\nu_{V}^\mathrm{sp}}{-E_{0B}(\epsilon,\gamma,\xi_{ij})+V}
~\partial_{\xi_{ij}} V}
{\sum_{V}\frac{\nu_{V}^\mathrm{sp}}{-E_{0B}(\epsilon,\gamma,\xi_{ij})+V}} 
\right|_{\xi_{ij}=0},
\end{eqnarray}
where $E_{0B}(\epsilon,\gamma,\xi_{ij})$ is the ground-state 
energy of the Hamiltonian (\ref{Hij}) and is calculated 
as explained in Section \ref{bosons}.

\section{Numerical results}
\label{numericalresults}
Now we apply the approach developed in the previous Sections to some
example cases. In particular, we compare the ground-state energy
obtained by Eqs. (\ref{E3}) and (\ref{E1F}) with that from
exact numerical calculations.

In our approach the starting point is the evaluation of the
statistical moments $\overline{\bm{\nu}}$ and $\bm{\Sigma}$.
These are obtained by generating trajectories in the lattice 
configuration space and counting the multiplicities $N_V$ and $N_A$.
The length of the trajectories is chosen to be sufficiently large 
for the asymptotic behavior to be established.
The determination of $\overline{\bm{\nu}}$ and $\bm{\Sigma}$
with good statistical precision requires a number of trajectories
which increases no more than linearly with the number of lattice sites 
$|\Lambda|$.
Therefore the evaluation of these moments is feasible even for large systems. 
In the following applications the statistical errors associated to
the measurement of $\overline{\bm{\nu}}$ and $\bm{\Sigma}$
are negligible on the scales considered.

\begin{figure}[t]
\centering
\psfrag{2x3 2+2}[][][0.8]{$2 \times 3$ $N_p=2$}
\psfrag{2x3 3+3}[][][0.8]{$2 \times 3$ $N_p=3$}
\psfrag{4x4 8+8}[][][0.8]{$4 \times 4$ $N_p=8$}
\psfrag{4x4 5+5}[][][0.8]{$4 \times 4$ $N_p=5$}
\psfrag{x}[t][]{$\gamma/\epsilon$}
\psfrag{y}[b][]{$E_{0B}/(\epsilon N_p)$}
\includegraphics[width=0.6\columnwidth,clip]{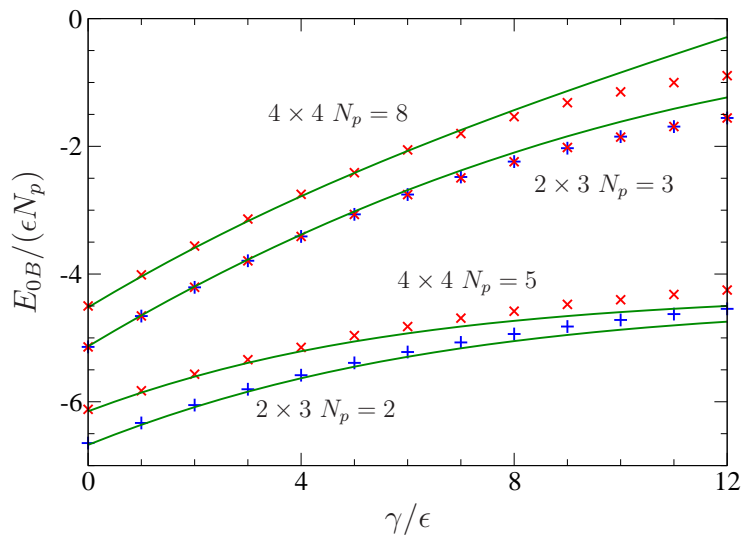}
\caption{Ground-state energy per particle 
for the hard-core boson FNU model
vs the interaction strength $\gamma$. 
The results from Eq. (\ref{E3}) (solid lines) are compared
with those from exact diagonalizations (\textcolor{blue}{$\bm+$}) 
and QMC simulations (\textcolor{red}{$\bm\times$})
for two-dimensional systems with periodic boundary conditions
and different $L_x \times L_y$ sites and $N_\ua=N_\da=N_p$ particles.
The statistical errors in the QMC data are negligible in this scale.}
\label{gauss.bose.eps}
\end{figure}

In Fig. \ref{gauss.bose.eps} we show the behavior of $E_{0B}$
for the hard-core boson FNU model 
as a function of the interaction strength $\gamma$ for several lattice 
systems. 
The solution of Eq. (\ref{E3}) compares rather well with the results 
of exact diagonalizations or QMC simulations. 
There is a small systematic error which grows with increasing $\gamma$
and/or the system size and which is maximum at half density. 
This error is related to the Gaussian shape (\ref{GAUSSIAN}) 
assumed for the asymptotic probability density.
In fact, the mentioned central limit theorem for
Markov chains applies to the variables $\nu_{V}\sqrt{N}$ and
$\nu_{A}\sqrt{N}$ while the function $g_N$ given by Eq. (\ref{G}) 
depends on $\nu_{V}N$ and $\nu_{A}N$.
This implies that the tails of the probability density give 
a finite contribution to the integral (\ref{AVERAGE2}).
From the structure of $g_N$, furthermore, it is evident that this error 
becomes large when the components of the vector $\bm{u}$ assume large values,
\textit{i.e.} for large values of $\gamma$ and/or large system sizes.

\begin{figure}[t]
\centering
\psfrag{2x3 2+2}[][][0.8]{$2 \times 3$ $N_p=2$}
\psfrag{2x3 3+3}[][][0.8]{$2 \times 3$ $N_p=3$}
\psfrag{4x4 8+8}[][][0.8]{$4 \times 4$ $N_p=8$}
\psfrag{4x4 5+5}[][][0.8]{$4 \times 4$ $N_p=5$}
\psfrag{x}[t][]{$\gamma/\epsilon$}
\psfrag{y}[b][]{$E_{0F}/(\epsilon N_p)$}
\includegraphics[width=0.6\columnwidth,clip]{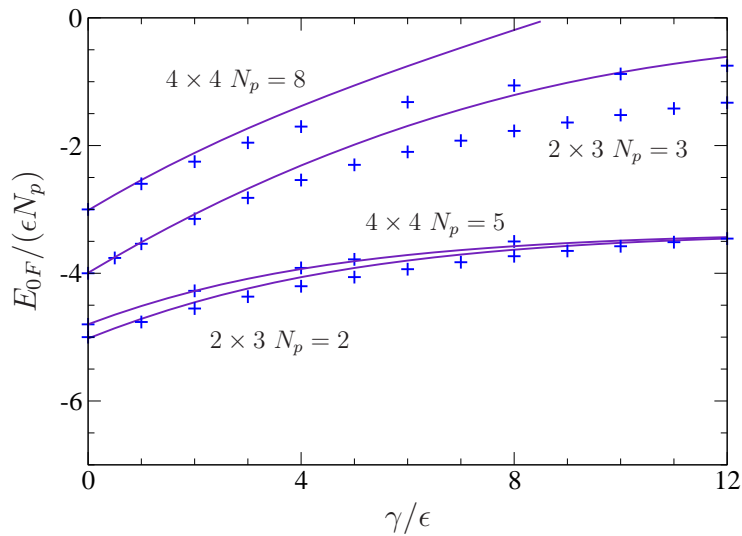}
\caption{
As in Fig. \ref{gauss.bose.eps} for the fermion FNU model. 
In this case no QMC simulations are available. 
Exact diagonalization data for the $4 \times 4$ systems 
are taken from \cite{FOP} ($N_p=8$) and 
\cite{PSBCPT,HFM} ($N_p=5$).}
\label{gauss.fermi.eps}
\end{figure}

In Fig. \ref{gauss.fermi.eps} we show the behavior of $E_{0F}$
evaluated according to Eq. (\ref{E1F})
as a function of the interaction strength $\gamma$ 
for the same cases considered in Fig. \ref{gauss.bose.eps}.
Compared with the hard-core boson case we observe a further
systematic error due to the approximation $s_N(\gamma) \simeq s_N(0)$.
Depending on the particle density of the system, this error 
adds to or subtracts from the systematic error due to the 
Gaussian tails of the probability density 
discussed in the case of hard-core bosons. 

We stress that, once $\overline{\bm{\nu}}$ and $\bm{\Sigma}$ are known,
our approach provides any other ground-state quantity analytically 
as a function of the Hamiltonian parameters.
On the other hand, QMC methods require, 
due to the unavoidable branching or reconfiguration 
techniques \cite{CALANDRASORELLA}, 
different simulations for different values of the parameters.

\begin{figure}
\centering
\psfrag{xx}[][][0.8]{$\gamma/\epsilon$}
\psfrag{yy}[][][0.8]{$\langle E_0 | S_i S_j | E_0 \rangle$}
\psfrag{x}[t][][0.8]{$q_x$}
\psfrag{y}[tl][][0.8]{$q_y$}
\psfrag{z}[r][][0.8]{$S(q_x,q_y)$}
\psfrag{g0}[][][0.8]{$\gamma=0$}
\psfrag{d0}[][][0.8]{$d_{ij}=0$}
\psfrag{d1}[][][0.8]{$d_{ij}=1$}
\psfrag{d2}[][][0.8]{$d_{ij}=2$}
\includegraphics[width=0.9\columnwidth,clip]{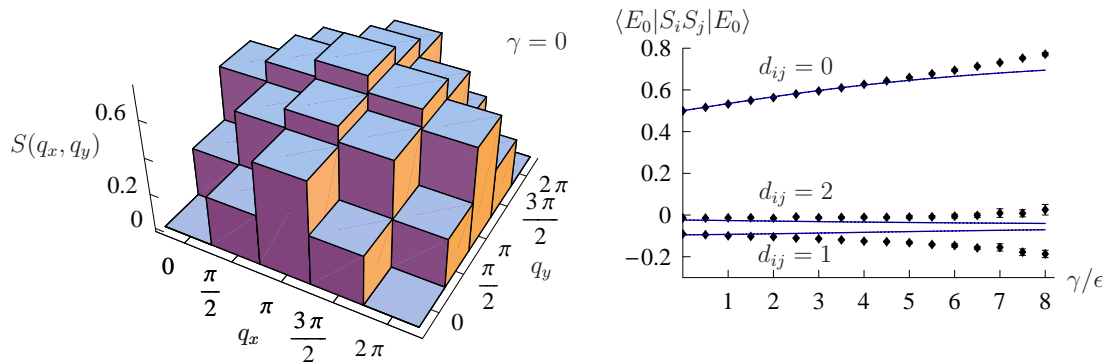}
\caption{Spin-spin structure factor $S(q_x,q_y)$
for the hard-core boson FNU model at $\gamma=0$
in a lattice with $4 \times 4$ sites, $N_\ua=N_\da=8$ particles,
and periodic boundary conditions.
For the same system, in the right plot we show the values of 
the ground-state expectation of the operators $S_iS_j$ for $d_{ij}=0,1,2$
as a function of the interaction strength $\gamma$ (solid lines)
compared with Monte Carlo results (dots with error bars).} 
\label{gauss.bose.structurefactor.eps}
\end{figure}

As an example of calculation of correlation functions in the ground state,
in Fig. \ref{gauss.bose.structurefactor.eps} 
we report the spin-spin structure factor $S(q_x,q_y)$
evaluated by using Eqs. (\ref{S}-\ref{S1})
for the hard-core boson FNU model. 
The value $S(\pi,\pi)$ represents a maximum for $S(q_x,q_y)$ and
for large values of $|\Lambda|$ is related to the staggered magnetization 
$m_s$ through $m_s=\sqrt{S(\pi,\pi)/ |\Lambda|}$.
Note that in the $4\times 4$ system with periodic boundary conditions 
considered in Fig. \ref{gauss.bose.structurefactor.eps} we have
only five different values for $\langle E_0 | S_i S_j | E_0 \rangle$
corresponding to the five possible distances 
$d_{ij}= | x_i - x_j | + |y_i - y_j|=0,1,2,3,4$. 
In Fig. \ref{gauss.bose.structurefactor.eps} we also show the
behavior of $\langle E_0 | S_i S_j | E_0 \rangle$ 
as a function of $\gamma$ for $d_{ij}=0,1,2$.
These terms provide the most important contributions to $S(\pi,\pi)$. 
For small values of $\gamma$, the results compare rather well with 
Monte Carlo data.

\section{Conclusions}
\label{conclusions}
By using saddle-point techniques and a central limit theorem, 
we have exploited an exact probabilistic representation of the
quantum dynamics in a lattice to derive analytical approximations 
for the matrix elements of the evolution operator in the
limit of long times. 
For both hard-core boson and fermion systems, 
this development yields to a simple scalar equation
for the ground-state energy.
This equation depends on the values of the generalized potentials $V$
and of the kinetic quantities $T$, and on the statistical moments 
$\overline{\bm{\nu}}$ and $\bm{\Sigma}$ of their asymptotic 
multiplicities $N_V$ and $N_T$.
In turn, these moments depend only on the structure of the system
Hamiltonian, not on the values of the Hamiltonian parameters.
This implies that the statistical moments must be
measured \textit{una tantum} for a given Hamiltonian structure and, 
once $\overline{\bm{\nu}}$ and $\bm{\Sigma}$ are known, 
our approach provides the ground-state energy analytically 
as a function of the Hamiltonian parameters.

In the long time limit, 
the saddle-point integrations used in our approach 
are asymptotically exact
and the central limit theorem evoked applies rigorously
to the rescaled multiplicities $N_V/\sqrt{N}$ and $N_T/\sqrt{N}$.
However, since functions depending on $N_V$ and $N_T$ are involved,
we have a small systematic error related to 
the finite contributions from the tails of the probability density.
This systematic error could be reduced by a large deviation analysis.
In fact, equations (\ref{E_{0B}0}) and (\ref{E3}) suggest 
that the Gaussian approximation for the probability density 
corresponds to a second order truncation of a cumulant expansion.
Anyway, the present Gaussian approach has the following 
relevant features:
\textit{i)} the equations derived for the ground-state energy 
and the ground-state correlation functions are particularly simple;
\textit{ii)} the corresponding results compare rather well to 
the exact ones in regions of physical interest.

In this paper we have considered two-dimensional lattice models 
at imaginary times.
Our approach, however, is valid in any dimension. 
Similar analytical expressions can be obtained also for the real 
time evolution.

\section*{Acknowledgments}
We thank F. Cesi for stimulating discussions
and G. Jona-Lasinio for insightful comments
and a critical reading of the manuscript.
This work was supported in part by
Cofinanziamento MIUR protocollo 2002027798\_001.

\end{document}